\documentclass{aa}
\usepackage[dvips]{graphicx}
\usepackage[dvips]{epsfig}
\newcommand{\ZPEG}{{\it Z-PEG}}
\newcommand{\zspe}{z_{\textrm{\tiny spe}}}
\newcommand{\zphot}{z_{\textrm{\tiny phot}}}
\newcommand{\zZPEG}{z_{\textrm{\tiny ZPEG}}}
\newcommand{\zFS}{z_{\textrm{\tiny FS}}}
\newcommand{\zfor}{z_{\textrm{\tiny for}}}
\newcommand{\deltaz}{\overline{\Delta z}}
\newcommand{\sigmaz}{\sigma_z}
\newcommand{\ageu}{age$_{\textrm{\tiny U}}$}

\begin{document}
\title{Photometric redshifts from evolutionary synthesis with P\'EGASE: the code \ZPEG~and the $z=0$~age constraint}
\authorrunning{D. Le Borgne \& B. Rocca-Volmerange} 
\titlerunning{Photometric Redshifts with P\'EGASE}

\author{Damien Le Borgne\inst{1} \and Brigitte Rocca-Volmerange\inst{1,2} } 
\offprints{leborgne@iap.fr}
\institute{Institut d'Astrophysique de Paris, 98\,{\it bis},
Boulevard Arago, F-75014 Paris, France 
\and
Institut d'Astrophysique Spatiale, B\^at. 121, Universit\'e Paris XI,
F-91405 Orsay, France}

\date{Received 19 November 2001 / Accepted 8 February 2002}

\abstract{Photometric  redshifts are  estimated  on the  basis of  
template scenarios with the help of the code \ZPEG, an extension of the
galaxy evolution  model P\'EGASE.2 and available on
the P\'EGASE web site. The spectral energy distribution (SED) templates are
computed  for  nine  spectral  types including  starburst,  irregular,
spiral and  elliptical.  Dust, extinction  and metal effects
are  coherently taken  into  account, depending on evolution scenarios.  
 The  sensitivity   of  results   to adding 
near-infrared  colors and  IGM absorption  is analyzed.   A comparison
with  results  of  other   models without evolution measures the  evolution  factor which
systematically increases the  estimated photometric redshift values by
$\Delta z \geq 0.2$ for $z > 1.5$.  Moreover we systematically check that
the evolution scenarios match  observational standard templates 
of nearby galaxies, implying an  age  constraint of the  stellar  
population at $z=0$ for each type.  The  respect  of this
constraint makes it possible to significantly  improve the accuracy of photometric
redshifts by  decreasing  the  well-known degeneracy  problem.   The
method  is applied  to the  HDF-N sample  (see in  Fern\'andez-Soto et
al.,1999).  From fits on SED templates by a $\chi^2$-minimization procedure, not
only is the  photometric redshift  derived but also  the corresponding
spectral type and the formation redshift $z_{\textrm{for}}$ when  stars first formed.  
Early epochs
of galaxy formation $z > 5$~are found from this new method and results
are compared to  faint galaxy count interpretations.  The  new tool is
available at: \texttt{http://www.iap.fr/pegase}
}

\maketitle
\section{Introduction}
The  determination of  galaxy distances  is  so crucial  for clues  on
galaxy evolution and cosmic structures that a large variety of methods
is currently being explored.  Spectroscopic determinations  are the most
precise but  they consume excessive observing time for  deeper and deeper
large   galaxy   samples.    For instance,   the   redshift   surveys
such as CFRS(\cite{CFRS}),  2dF \cite{Folkes}), Hawaii(\cite{Cowie})  and more
recently the SLOAN, with millions  of targets with various spectral types,
are complete to  $z \leq 1.5$.  At higher redshifts $z > 1.5$, the 
galaxy populations observed at  faint magnitudes in deep surveys cover
a large range  of redshifts which will be  easily accessible from
photometry. However  many problems  of degeneracy,  number and  
width of filters and extinction first have  to  be clarified.  
Typical  SED
features like the 4000\AA~discontinuity or  the Lyman break are known to
be  fruitful  signatures  for evaluation of redshifts when compared with
template SEDs.   \cite{Steidel} proposed an empirical  method based on
these discontinuities  to detect $z  \geq 4$ galaxies.  Successful in
discovering  distant  sources,  the   method  is  however  imprecise  and
prone to degeneracies.  The comparison of observed  SEDs with calibrated
templates on an  extended wavelength range is the  best way to rapidly
determine  redshifts  of  a  large  number of  faint  galaxies,  on  a
continuous range $0  \leq z \leq 4$ .   Such comparisons were proposed
with  templates  from  no-evolution models  by \cite{Baum},  \cite{Koo},
\cite{Loh} and more recently Fern\'andez-Soto et al. (1999) (hereafter
FSLY); others  proposed evolutionary SED  methods such as \cite{Bolzonella} and
\cite{Massarotti} using templates derived from a variety of evolutionary
codes.   However the  evolutionary  codes and  their applications  may
differ.  If results  are roughly similar at low redshifts 
(see \cite{Crete}), they may actually strongly differ from each other at
high  redshifts, depending  on  adopted star  formation  laws and  the
corresponding age  constraints, initial  mass function, dust  and metal
effects as well as interpolation algorithms.

An essential property  of most codes is the  large wavelength coverage
from  the   far-UV  to  the  near-infrared  needed to   compute  SEDs that are highly
redshifted.   Moreover  our  code  P\'EGASE.2, \cite{FRV1},  in 
its current version (see next footnote)  is able  to  take  into
account  metallicity effects  in its  stellar library  and isochrones.
Evolution scenarios  of nine spectral types, defined  by star formation
parameters, have  been selected to reproduce  the observed statistical
SEDs of  $z=0$~ galaxy templates, \cite{FRV99stats}.  Then two correction
factors (cosmological k-correction and evolutionary e-correction) are
computed with the model to predict redshifted SEDs, in order to be used as
comparison  templates to observations. Other PEGASE.2 scenarios based on  
different prescriptions of dust properties or star formation parameters might be 
computed and used as templates only if they respect fits of $z \simeq 0 $ 
galaxy properties. This is beyond the objectives of this paper.
 Another method  of photometric
redshift determination supposes that the evolution effect is dominated by
shot noise, \cite{Connolly}.

We   present in  section  2 a  new  tool, \ZPEG, to  estimate
photometric redshifts  on a continuous  redshift range $0 \leq  z \leq
6$~or more.  Observed colors  or spectra are statistically compared to
the P\'EGASE.2  atlas for 9 spectral types  (starburst: SB, irregular:
Im, spiral:  Sd, Sc,  Sbc, Sb, Sa,  SO and elliptical:  Ell galaxies).
Section 3 presents the results  for the well-known Hubble Deep Field North
and  the comparison  with spectroscopic  redshifts allows us to derive average
values of evolution factors.  Section 4 shows the  sensitivity to various
parameters   such as   the  NIR   colors and IGM  absorption. Interesting
consequences  for the   redshifts of formation of the  sample are
derived in section 5.  Discussion and conclusions are  proposed
in sections 6 and 7 respectively.
\section {The code \ZPEG~(web available)}
\subsection {The evolution scenarios of galaxies}
The atlas  of synthetic  galaxies used as  templates is  computed with
 P\'EGASE.2 on the basis of evolution scenarios of star formation. The
 synthesis method assumes that  distant galaxies are similar to nearby
 galaxies, but look  younger at high $z$ since  they are seen at
 more remote epochs.  Respecting  this constraint will make our
 redshift  determinations more robust.   In   an  earlier  paper (\cite{FRV1}),  we
 investigated the selection of  star formation rates able to reproduce
 the   multi spectral   stellar    energy   distributions   of   nearby
 galaxies.   Another    article by \cite{FRV99stats}   computed   the
 statistical  SEDs of  about  800 nearby  galaxies  observed from  the
 optical and  the near-infrared for  eight spectral types  of galaxies
 and used to  fit scenarios at $z$=0. Star  formation rates (SFRs) are
 proportional  to  the  gas   density (with one exception, see table 
\ref{table:scenarii}), the astration   rate  $\nu$~increasing  from 
 irregular  Im to  elliptical E  galaxies. The
 star formation rate of starburst SB scenario is instantaneous. Infall
 and   galactic  winds   are  typical   gaseous  exchanges   with  the
 interstellar  medium. They  aim  to  simulate the  mass
 growth and  to subtract the  gas fraction 
 (preventing any further star formation) respectively. 
Unlike other studies (\cite{FSLY2001a}, \cite{Massarottib}), our scenarios
do not need to add any starburst component to be consistent with $z \simeq 0$
observations.\\
\begin{table}
\caption{{\small PEGASE.2 scenarios used as template parameters. SFR=$\nu \times M_{\textrm{gas}}$, 
except for starbursts and irregular galaxies. $\nu$~ is in units of $\textrm{Gyr}^{-1}$~ and  M$_{\textrm{gas}}$~
is the gas density. Infall time-scales are  in Myrs. 
The dust distribution is fitted on a King profile for E and S0, 
while an inclinaison-averaged disk distribution  is applied to spiral and irregular galaxies
(see text for details). 
Starburst galaxies have no extinction correction. For all the scenarios, the age of the universe is an upper limit on the age.}}
\begin{center}
\begin{minipage}{8cm}\def\footnoterule{}
\begin{center}
\begin{tabular}{lllll}
Type   	& $\nu$     &infall  &gal. winds &  age at $z=0$\\
\hline
\hline
SB 	&$\delta(t)$	&		&          & 1~Myr to 2~Gyr   \\
E   	&3.33		&300		& 1 Gyr    & $>13$~Gyr\\
S0	&2		&100		& 5 Gyr    & $>13$~Gyr\\
Sa	&0.71		&2\,800		&          & $>13$~Gyr\\
Sb	&0.4		&3\,500		&          & $>13$~Gyr\\
Sbc	&0.175		&6\,000		&          & $>13$~Gyr\\
Sc	&0.1		&8\,000		&          & $>13$~Gyr\\
Sd	&0.07		&8\,000		&          & $>13$~Gyr\\
Im	&0.065\footnote{for this scenario only , we have SFR=$\nu \times M_{\textrm{gas}}^{1.5}$}	&8\,000		&		& $>9$~Gyr \\
\end{tabular}
\end{center}
\end{minipage}
\label{table:scenarii}
\end{center}
\end{table}
The initial mass function (IMF) (\cite{Rana}), is used in our evolution
scenarios. However \cite{Giallongo} showed  that the choice of the IMF
does not influence much  the photometric redshift estimates of high-$z$~
candidates   ($z>2.5$). \\
P\'EGASE.2 is the most
recent version of P\'EGASE, available by ftp and on a web 
site\footnote{at anonymous ftp: \texttt{ftp.iap.fr/pub/from\_users/pegase}
and \texttt{http://www.iap.fr/pegase}}.
Non-solar metallicities are implemented in stellar tracks and spectra 
but also a far-UV spectral library for
hot stars (\cite{Clegg}) complements the \cite{Lejeuneetal} 
library. The  metal   enrichment  is followed  through
the successive generations of stars and is taken into account for spectra
of the  stellar library as well  as for isochrones.   
In P\'EGASE.2, a consistent treatment of the
internal extinction  is proposed by fitting  the dust amount
on metal abundances. The  extinction factor depends  on the
respective spatial  distribution   of  dust   and  stars  as   well  as   on  its
composition. Two patterns are modeled with either the geometry of bulges
 for elliptical galaxies or disks for spiral galaxies. In elliptical galaxies, the
dust distribution  follows a King's profile. The density  of dust is described
as a power of  the density of stars (see \cite {FRV1}  for details). Through
such a geometry, light  scattering  by   dust  is   computed  using   a  transfer
model, outputs of which are  tabulated in  one  input-data file  of the  model
P\'EGASE.  For spirals and  irregulars, dust is distributed along a
uniform plane-parallel slab and mixed with gas. As  a direct consequence, 
 the synthetic templates  used to
determine  photometric redshifts at  any $z$, as well  as to  fit the
observational   standards  at   $z=0$,  are   systematically  reddened.
  We  also  add  the  IGM
absorption  following  \cite{Madau95}  on  the  hypothesis  of  
Ly$_\alpha$, Ly$_\beta$, Ly$_\gamma$ and Ly$_\delta$ line
blanketing  induced  by \ion{H}{i}  clouds, 
Poisson-distributed along the line of sight.
This line blanketing can be expressed for each order of the Lyman series by 
an effective optical depth $\tau_{\textrm{\small eff}}=A_i \times (\lambda_{\textrm{\small obs}}/\lambda_i)^{1+\gamma}$, 
with $\gamma=2.46$ and $\lambda_i=1216, 1026, 973, 950$ \AA~ for Ly$_\alpha$, Ly$_\beta$, Ly$_\gamma$ and Ly$_\delta$ respectively.
The values of $A_i$  are taken from \cite{Madau96}, in agreement 
with the \cite{Press93} analysis on a sample of 29 quasars at $z>3$.
We shall see below  that the IGM absorption  alters the visible and
IR colors  more than about $0.1$~mag  as soon as $z >2 $, leading to
a  more  accurate  determination  of photometric  redshifts  at  these
distances.\\
For each spectral
 type,  a typical  age of  the  stellar population  is derived.   Time
 scales, characteristics and ages of the scenarios  are  listed in  table
 \ref{table:scenarii}.
\subsection {The $\chi^2$ minimization procedure}
\label{section:chi2min}
A  3D-subspace  of  parameters  (age, redshift, type) is defined  by  the
template sets. It is used  to automatically fit   observational data.  This
subspace  in the age--redshift plane is limited by  the cosmology in
order to avoid inconsistencies: a 10 Gyr old galaxy at $z=2$~ cannot
exist    in   the    standard   cosmology\footnote{we    assume 
H$_0$=65 km.s$^{-1}$.Mpc$^{-1}$;~$\Omega_M=0.3$~;~$\Omega_\Lambda=0.7$}   because   at  this
redshift, the age of the universe is about 5 Gyr. Moreover the subspace
is also
limited  by  the  age  (redshift  corrected) imposed  by  the  adopted
scenario of spectral type evolution.  As an example, if elliptical and
spiral galaxies  must be at  least 13 Gyr  old at $z=0$, it  means at
least 5 Gyr  at $z=1$ and so on.\\ 
Each point  is granted a synthetic
spectrum; its flux through the filter $i$ is called 
$F^{synth}_i$. For each point of this 3D-subspace, the
fourth parameter $\alpha$ is  computed with a $\chi^2$ minimization to
fit as  well as possible  the observed fluxes in filters:\\
\begin{equation}
\chi^2= \sum_{i=1}^{\it N}\left[\frac{F^{obs}_i-\alpha \times F^{synth}_i}{\sigma_i}\right]^2
\end{equation}

$N$~is  the  number  of  filters,  $F^{obs}_i$~  and  $\sigma_i$~  are
 the  observed flux and  its error bar through  the filter
$i$ respectively. In the case of an observed spectrum without redshift signatures,  the sum can also be computed from wavelength bins.
Then, each point of the  3D-subspace of parameters has a $\chi^2$
value.   A  projection  of  this  3D-$\chi^2$ array  on  the  redshift
dimension     gives     the    photometric    redshift    value
$\zphot$.

The values of $\sigma_i^2$ can be evaluated by the quadratic sum 
of the systematic errors and of the statistical errors. The extremely low
values of observational errors, adopted as statistical, may result in
anomalously high reduced $\chi^2$  minima.  
In this study we consider as negligible systematic errors,
keeping in mind that it maximizes the $\chi^2$ minimum value 
(possibly up to 100).  
In such a  case,  statistical rules
claim that the  result (the photometric redshift) is  not reliable and
has to  be excluded.   Yet, with such prescriptions,  most  of the  results
 would be excluded,  because the  photometric  errors of  the
observations are very  low.  This is why all  the primary
solutions are often kept, including cases of  very high reduced $\chi^2$ minima. In 
the following, we will also adopt this philosophy. However, our error bars 
might appear larger than in the previous studies, that limit
their results to one unique but less robust solution. Indeed,
the estimation of  the error bar of  a photometric redshift  is often
estimated  by the  redshifts for  which $\chi^2  \leq \chi^2_{\textrm{
\tiny min}}+1$. This method is only valid when the minimum reduced
$\chi_{r,\textrm{\tiny min}}^2  \simeq 1$ (otherwise the  error bar is
very  underestimated).\\ 
We  choose to  estimate the  error bar  by the
redshift  values  for  which  $\chi_n^2  \leq  \chi^2_{n,\textrm{\tiny
min}}+1$,  where  $\chi_n^2$  is  the  $\chi^2$  ``normalized''  with
$\chi^2_{r,\textrm{\tiny min}}=1$ .  The error bar is then much larger
and  may lead  to  secondary solutions.  \cite{FSLY2001b} use another accurate estimation
of  the error  bars  which  also gives  secondary  solutions, for  the
$3~\sigma$ level for instance. This  is the  case when the Lyman and Balmer breaks are hardly 
distinguished, as an example.   
\subsection {The \ZPEG~interface}
\ZPEG~is an interface available on the P\'EGASE web site (see footnote on page 2).
Inputs  are  fluxes or  colors  of  the  observed galaxies and their error bars. Sets
of classical
filters defined in a variety  of photometric systems are proposed. It is
possible  to use  a  user-defined  filter,  with   its  passband  and
calibration.  In the general case, for which no information on
the galaxy  spectral type  is {\it a  priori} known,  the minimization
procedure is  tested on  all the template  types.  It is also possible to
restrain the template galaxies to a given type. Types are then chosen among the
9  pre-computed synthetic  spectra. The  default redshift  sampling is
$0.25$ and  may be reduced on  a narrower redshift range.  The age and
redshift axes  are respectively defined  by step - up  and - down limits.
The user  will choose the cosmological  parameters ($H_0$, $\Omega_M$,
$\Omega_\Lambda$), on  which the  time-redshift relation depends.  The
default  values of cosmological  parameters are  those adopted  in this
 article.

Outputs are the estimated photometric redshift $\zphot$~and its
error bar, the age and the spectral type of the best fitted synthetic galaxy. 
A $\chi^2$ projection map in the age--redshift plane and
its projection on the redshift and age axes (see Fig.~\ref{figure:chi2_z_t}) are also presented.
\subsection{The coherency test}
\begin{center}
\begin{figure}[!htbf]
\begin{center}
\includegraphics{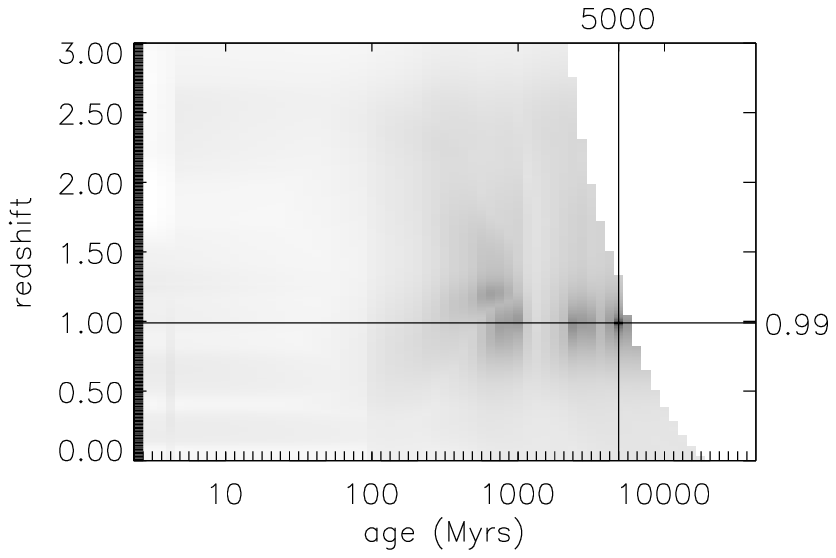}
\includegraphics{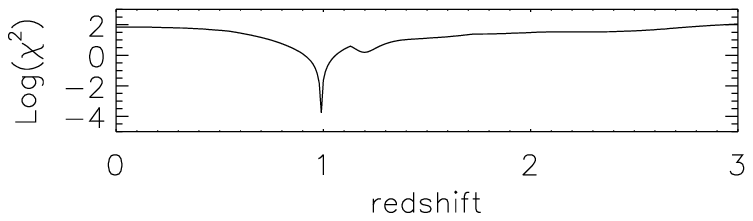}
\end{center}
\label{figure:proj}
\caption{{\small  The test of coherency on the $z=0.989$ and $5$ Gyr old galaxy: 
darker grey zones correspond to lower $\chi^2$ values in the age -- z plane.
The minimum $\chi^2$ location is the cross center of the two solid lines.
The plot log$_{\textrm{\tiny 10}}$($\chi^2$)  versus redshift estimates the accuracy of the minimum
and shows possible secondary minima. }}
\label{figure:chi2_z_t}
\end{figure}
\end{center}

The relevance of the fit procedure is checked on a synthetic galaxy of the atlas
with a  known $z$. 
As an example, input data are the colors computed from the SED of the Sbc galaxy 
at age 5 Gyr and  redshift $z=0.989$. The output value is, 
as expected, $z=0.99$, identical to the input galaxy redshift ($z=0.989$)
with a precision of $\Delta z=0.01$ (no photometric error). 
Figure \ref{figure:chi2_z_t} shows the resulting projected $\chi^2$ map on the 
age $vs$~redshift plane for this  test.
\section {Photometric redshifts with evolution}
\subsection{The HDF-N galaxy sample}
The Hubble Deep Field North (HDF-N) catalog (\cite{Williams})
was chosen to test the evolution factor. Spectroscopic redshifts
were measured carefully by \cite{Cohen2000}, with some additions and 
corrections in \cite{Cohen2001}. \cite{FSLY} give photometric 
data for the HDF-N objects and \cite{FSLY2001a} give a 
correspondence between their objects and the objects in \cite{Cohen2000}.
We exclude from the sample galaxies with negative  fluxes. 
The remaining  selected sub-sample contains 136 galaxies 
with redshifts distributed from $0$ to $5.6$.\\
The HDF-N sample  presents a series of advantages
to explore the accuracy of our method and to compare it
to others. With a large redshift range  
($0 \le z \le 5.6$), the sample data acquired with
the WFPC2/HST camera is one 
of the deepest (down to B=29) with an extent
over the wavelength range from about 3000\AA~to 8000\AA~ 
with the filters F300W, F450W, F606W, and F814W (hereafter called U,B,V,I). 
Moreover the near-infrared standard Johnson-Cousins J, H and K$_s$~(hereafter K) 
colors listed in \cite{FSLY} from \cite{Dickinson} were observed at the KPNO/4m telescope. 
The calibration used is AB magnitudes.
\[m_{\mathrm{AB}}=-2.5\log_{10}\frac{\displaystyle{\int F_{\nu}
\,T_{\nu}\,\mathrm{d}\nu}}{\displaystyle{\int T_{\nu}\,\mathrm{d}\nu}}-48.60\]
with $F_{\nu}$ in erg.s$^{-1}$.cm$^{-2}$.Hz$^{-1}$.
$T_{\nu}$ is the transmission of the filter.

A further advantage is the benefit of spectroscopic redshifts of the 
selected sub-sample,
as given by \cite{Cohen2001}. This allows a statistical comparison with our 
photometric redshifts and  an estimation of  the dispersion.
Another advantage of this sample is the possibility 
of measuring the type-dependent evolution factor 
by comparing our  results with
the \cite{FSLY}'s results, since the latter authors propose photometric redshift 
determinations based on a maximum
likelihood analysis and 4 spectral types without any evolution effect.
\subsection{The spectroscopic-$z$ -- photometric-$z$~ plane}
\begin{center}
\begin{figure}[!h]
\begin{center}
\includegraphics[width=9cm]{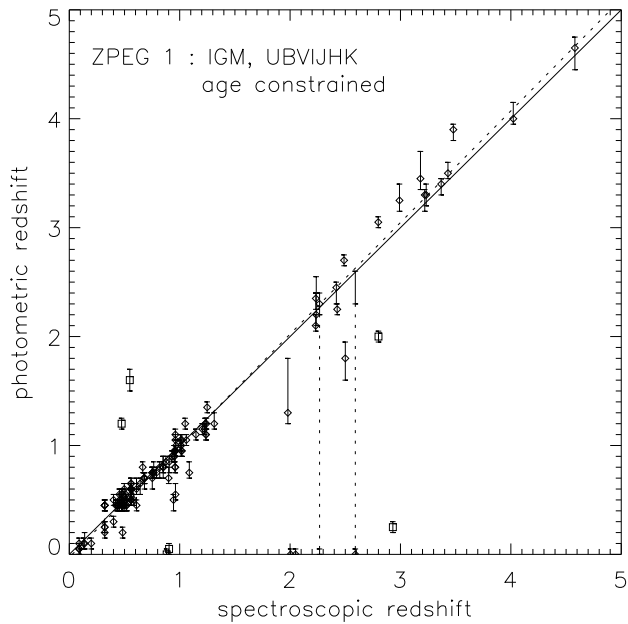}
\end{center}
\caption{{\small Comparison of photometric redshifts (points with error bars)
predicted by the model \ZPEG.1 to 
spectroscopic redshifts of the selected HDF-N sample. The solid line of slope=1 shows 
the case of equality
for comparison. The dotted line is the linear regression of our photometric redshifts 
 if we exclude the points with $|\zZPEG-\zspe| \geq 1$.
Predictions are computed
with IGM absorption from \cite{Madau96} and ISM reddening  according to P\'EGASE algorithms.
When two solutions or more are found (degeneracy), the error bars  
are linked by a dotted line. The squares are objects with discordant redshifts also pointed out by \cite{FSLY2001a}.}}
\label{figure:zspezphot}
\end{figure}
\end{center}

Figure \ref{figure:zspezphot} presents the plane $\zspe$ -- $\zphot$~ resulting
from our fit of the HDF-N sample with P\'EGASE templates.
\subsection{Evolution factors}
\subsubsection{Average evolution factor on $z_{\textrm{\small phot}}$}
\begin{center}
\begin{figure}[!htbf]
\begin{center}
\includegraphics{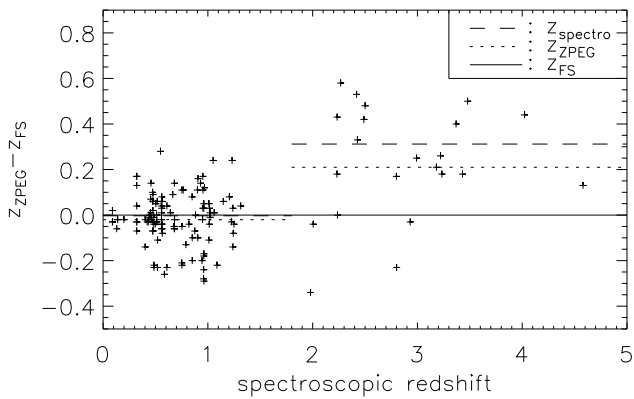}
\end{center}
\caption{{\small Crosses are differences 
$\zZPEG-\zFS$. For $\zspe > 1.5$, the evolution factor is measured by the distance between the 
dotted and full lines,
respectively the median value $<\zZPEG-\zFS>$
and 0. $\zFS$ are the \cite{FSLY}'s
determinations. The dashed line  is the median value $<\zspe-\zFS>$. 
For $\zspe<1.5$, the evolution factor has the same magnitude as the error bars.}}
\label{figure:figdeltaz}
\end{figure}
\end{center}

We compare our best results $\zZPEG$, from the evolutionary model \ZPEG.1 (selected from the variety
of models in table 2) to photometric redshifts $\zFS$
estimated 
by \cite{FSLY}. These authors only took into account
k-corrections of the galaxy spectral distributions from \cite{Coleman} while 
we simultaneously compute the k- and e- correction factors at any $z$.
Figure \ref{figure:figdeltaz} shows the difference for each galaxy of the sample.
Lines trace the median values of the difference of $\zFS$ with 
 \ZPEG.1, and $\zspe$~ values respectively. The zero value 
(difference to $\zFS$) is plotted for comparison. 
A systematic effect is observed for $z \geq  $1.5. The evolution effect is 
measured around $<\zZPEG-\zFS> = 0.2$, independent of type. 
The median\footnote{
note that the median operator is not transitive:\\ $<\zZPEG-\zspe> \neq <\zZPEG-\zFS>-<\zspe-\zFS>$.}
value  $<\zZPEG-\zspe> =- 0. 03$~
is measured and remains inside the error bars.
Evolution effects are more important for
distant galaxies and an appropriate evolutionary code is required for such estimates.
\subsubsection{Determination of galaxy spectral types}
\begin{center}
\begin{figure}[!htbf]
\begin{center}
\includegraphics{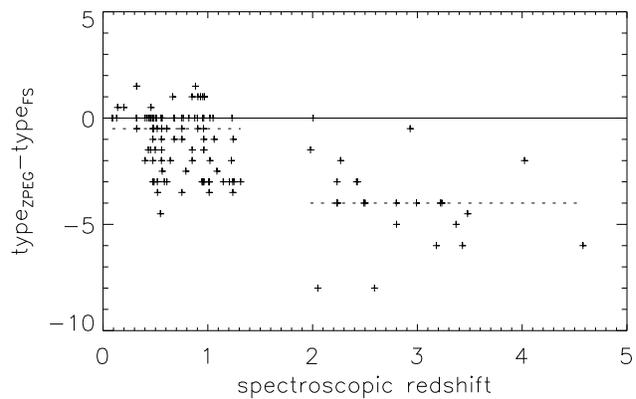}
\end{center}
\caption{{\small Comparison of our spectral type estimates to FSLY's.  The values are  
${\textrm type}_{\textrm{\it \tiny ZPEG}}-{\textrm type}_{\textrm{\tiny FS}}$ 
assuming the following values for the spectral types (type 0: starburst, 
1: elliptical, 2: S0, 3: Sa, 4: Sb, 5: Sbc, 6: Sc, 7: Sd, 8: Im)} }
\label{figure:typescomp}
\end{figure}
\end{center}

Our best fit gives a minimal $\chi^2$ for a triplet (age, redshift, type),
so that we can deduce the spectral type of the observed galaxies by comparing to the
SEDs of 9 different types.
The comparison of types derived by \ZPEG.1 and by \cite{FSLY} is
shown in Fig.~\ref{figure:typescomp}. In most cases, 'earlier type' galaxies 
are found with \ZPEG.1, particularly
 at high redshifts. This effect is expected from evolution scenarios   
because, when considering only the k-correction,  \cite{FSLY} would
find an evolved spiral galaxy when \ZPEG.1 finds a
young elliptical galaxy. A better definition of
the morphological properties  of galaxies in the near future will help us to 
arrive at a conclusion. 

\begin{table*}[!t]
\begin{center}
\caption {{\small The variety of  \ZPEG~models is tabulated with the 
adopted constraints
(IGM absorption, near-infrared JHK filters and age).
Statistical results (mean difference and dispersion) are given in last columns.
A comparison to \cite{FSLY} and \cite{Massarottib} results is also given. 
The $\deltaz$ and $\sigma$ are computed using primary and secondary solutions for \ZPEG~models (see Sect. \ref{section:chi2min}), 
and primary solutions only for FSLY and MIBV models.}}
\begin{minipage}{\textwidth}\def\footnoterule{}
\begin{center}
\begin{tabular}{|  l |  c c  c  c |r@{}r@{} l l|c|}
\hline 
\ZPEG~Model	& IGM abs.	&UBVI	&JHK	&age constraint & 	  &   &\multicolumn{1}{c}{$z<1.5$} 	&\multicolumn{1}{c|}{all $z$} 	& Figure\\
\hline          
\hline       
1		&x		&x	&x	&x		&$\deltaz$&$=$&$-0.0214$	&$ -0.0844$		&\ref{figure:zspezphot}\\  
		&		&	&	&		&$\sigmaz$&$=$&$\mathbf{0.0980}$&$\mathbf{0.4055}  $		&\\  
\hline															
2		&x		&x	&x	&		&$\deltaz$&$=$&$0.0251 $ 	&$ -0.0621$   		&\ref{figure:fig8}    \\    
		&		&	&	&		&$\sigmaz$&$=$&$\mathbf{0.1156}$&$\mathbf{0.4441}  $		&   \\    
\hline								  							
3		&		&x	&x	&		&$\deltaz$&$=$&$0.0252$ 	&$ -0.0589$		& \ref{figure:fig8}\\
		&		&	&	&		&$\sigmaz$&$=$&$0.1156$		&$ 0.5738$		&\\
\hline								  							
4		&x		&x	&	&		&$\deltaz$&$=$&$0.1273$ 	&$ -0.0040$		&\ref{figure:fignoJHK}\\    
		&		&	&	&		&$\sigmaz$&$=$&$0.3179$ 	&$ 0.5840$		&\\    
\hline
FSLY\footnote{results with \cite{FSLY}'s $\zphot$ on the same sample, for comparison, using only primary solutions}&x&x&x&	&$\deltaz$&$=$& $ -0.0037$	&$ -0.0579$	& \\
						      					& & & &					&$\sigmaz$&$=$& $ 0.1125$ 	&$ 0.2476$	& \\
\hline
MIBV\footnote{results from \cite{Massarottib} with P\'EGASE and additional starburst component on a similar sample, for comparison, using only primary solutions}
						         &x&x&x&	&$\deltaz$&$=$& $ 0.026$	&	& \\
						    	  & & & &	&$\sigmaz$&$=$&		$ 0.074$ 	&	& \\
\hline
\end{tabular}
\end{center}
\end{minipage}
\label{table:stats}
\end{center}
\end{table*}
\section{Sensitivity to parameters}
\label{section:parameters}
Table \ref{table:stats} summarizes the comparison $\zZPEG-\zspe$~
and the corresponding dispersion, given by a series of various \ZPEG~  models
computed by changing only one parameter. The considered parameters are age, absorption
by the IGM and the addition of NIR colors (JHK). 
We evaluate the offset and the dispersion of the 
$\zZPEG-\zspe$ distribution with 
\begin{equation}
\overline{\Delta z}=\frac{1}{N} \sum _1^N (\zZPEG-\zspe)
\end{equation}
\begin{equation}
\sigma^2_z=\frac{1}{N-1} \sum _1^N \left[ (\zZPEG-\zspe)-\overline{\Delta z}~\right]^2
\end{equation}

where N is the number of solutions. Most teams only limit their
results and corresponding error bars to primary solutions for $\zphot$. 
In such cases, N is the number of galaxies in the sample.
Yet, as the uncertainties on $\zphot$ are often underestimated (see Sect. \ref{section:chi2min}),
we choose to use $N=$number of primary solutions + number of secondary solutions. 
Thus, the estimation of the dispersion is made with more points than the number of galaxies.
As a consequence, the value of the dispersion differs from the dispersion computed using 
only primary solutions: the latter value might, by chance, be sometimes smaller. 
This is the case when the number of galaxies for which the primary solution 
is in agreement with the real redshift, dominates.
\subsection{Effect of the $z=0$~age constraint}
\begin{figure}[!htb]
\begin{center}
\begin{center}
\includegraphics{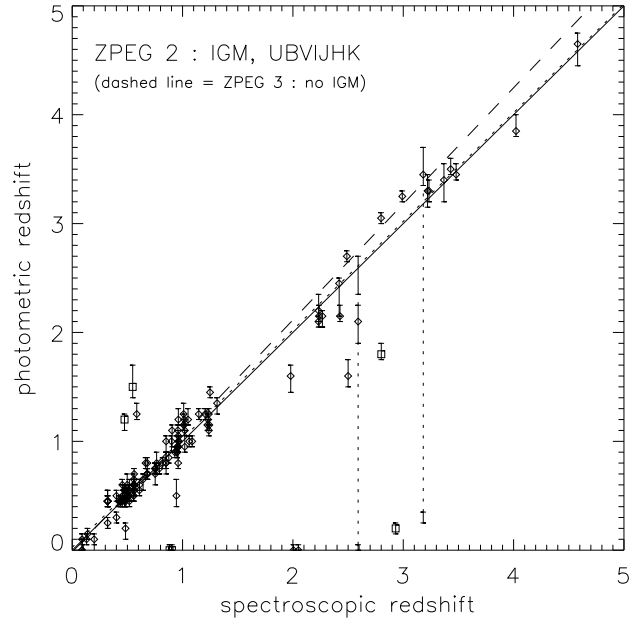}
\end{center}
\caption{{\small Photometric redshift estimations like in Fig.~\ref{figure:zspezphot} 
when our computations are free of any age constraint (model \ZPEG.2). 
The dotted diagonal line is the linear regression of points 
with $|\zZPEG-\zspe| < 1$. The dashed line 
is the linear regression without the IGM absorption (model \ZPEG.3)}}
\label{figure:fig8}
\end{center}
\end{figure}

\begin{center}
\begin{figure*}[!htb]
\begin{center}
\includegraphics{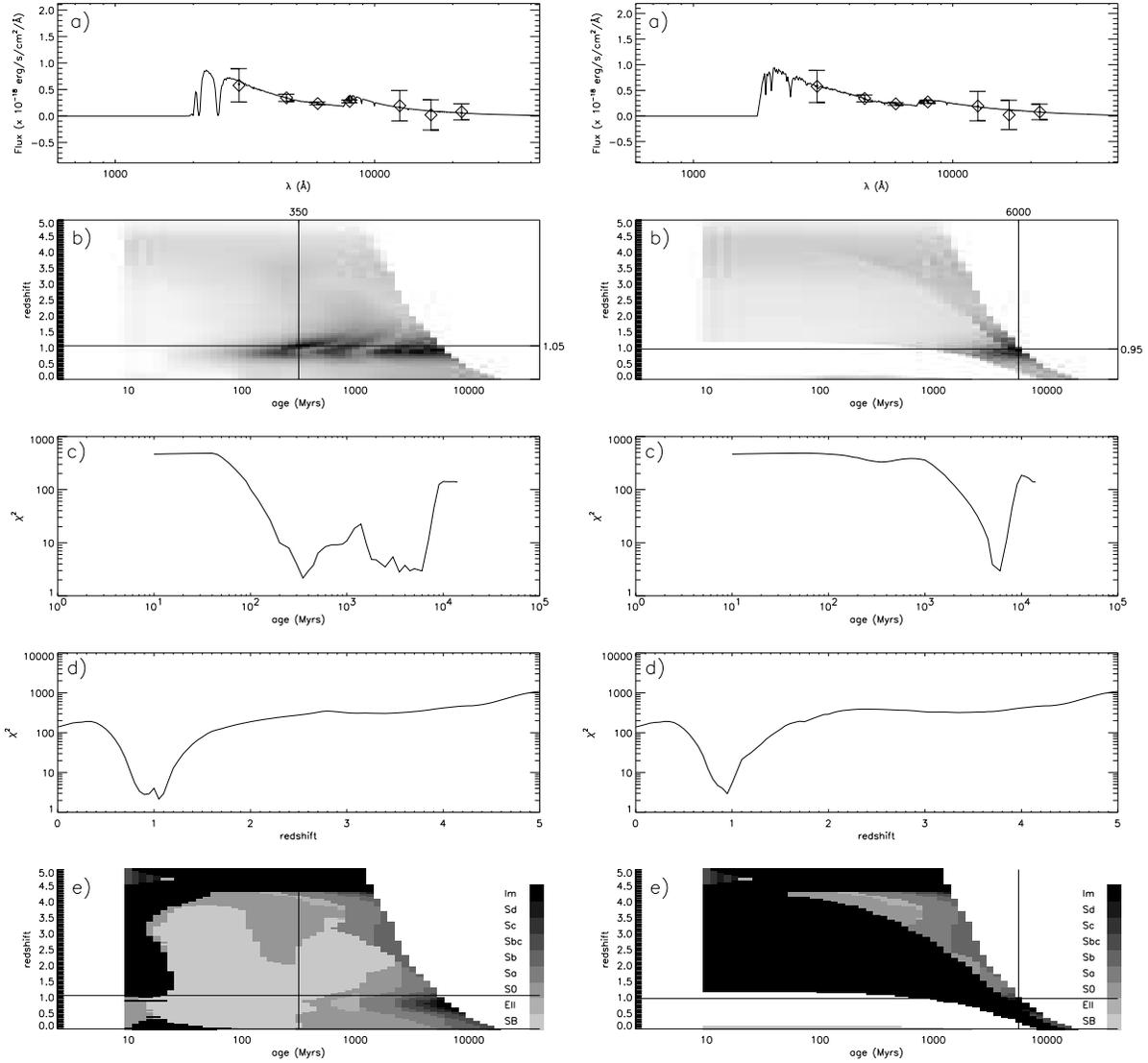}
\end{center}
\caption{{\small Fit of  galaxy  319 (number from \cite{FSLY}). 
On the left hand side, the fit is made with model  \ZPEG.2 
(without age constraint) and on the right hand side we use model \ZPEG.1 
(with age constraint): {\bf a)} observed colors with ten times magnified error bars (for plotting only) and best-fitting synthetic SED,
 {\bf b)} $\chi^2$ map on the age-$z$ plane (black corresponds to low $\chi^2$), 
{\bf c)} $\chi^2$ projection on the age axis, {\bf d)} $\chi^2$ projection on the redshift axis, 
and {\bf e)} Type maps on the age-$z$ plane; the age-$z$ solutions found in b) are also plotted. 
The model \ZPEG.2 (left) shows several minimum redshifts around $1$ whereas model 
\ZPEG.1 (right) only shows one minimum close to the spectroscopic redshift 
($z_{\textrm{spe}}=0.961$).}}
\label{figure:319}
\end{figure*}
\end{center}

The basic procedure of the spectral synthesis of distant galaxies
requires one to respect the observed SEDs of standard nearby galaxies.
The correction factors (k-correction for expansion and e-correction for
evolution) are computed  from the $z=0$~templates, fitting at best 
observational data.
As a consequence, a $z=0$~age of the 
stellar population is imposed by the synthetic SED template. 
When the so-called $z=0$~ age constraint is 
taken into account,  the dispersion
$\sigmaz$~
is reduced by 15\% for $z<1.5$ ($\sigmaz=0.098$  with model \ZPEG.1, Fig.~\ref{figure:zspezphot},
compared to $\sigmaz=0.116$  with model \ZPEG.2, Fig.~\ref{figure:fig8}). 
For $0<z<5$, the improvement factor is 8\%.
As discussed above, computations with evolution require the age 
constraint in order to make synthetic redshifted templates compatible 
with $z=0$ observed templates. As discussed below, a similar constraint is used
in the interpretation models of faint galaxy counts 
(see \cite{FRV99counts}). The remark is all the more important as
most photometric redshift models, even with evolution corrections, compute  redshifts
without the age constraint, finding fits with synthetic templates 
sometimes unable to reproduce $z=0$ galaxies.
Yet, we demonstrate here that our strong age constraint and the use of 
appropriate scenarios of evolution give better results than the often-used 
'no age constraint' method.\\
Moreover, not constraining the $z=0$~ages of galaxies (models \ZPEG.2 and beyond) 
increases the age--redshift--type degeneracy (see Fig.~\ref{figure:319}, 
left hand side): at a given redshift, a young ( $<$ 1Gyr) elliptical 
galaxy would be an acceptable
solution, with almost the same optical colors as an old irregular one. 
But using the model \ZPEG.1 
that takes care of ages, we restrain the acceptable range 
 in the age--redshift--type space and raise partially a degeneracy. 
Figure \ref{figure:319} clearly shows this effect on  galaxy number 319 of 
the \cite{FSLY} sample.
\subsection{Effect of the $\Delta \lambda$~ wavelength coverage}
\begin{center}
\begin{figure}[!htbf]
\begin{center}
\includegraphics{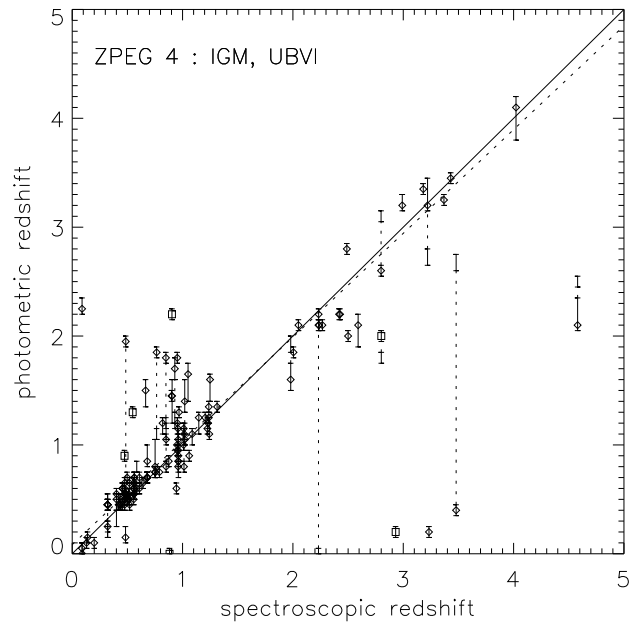}
\end{center}
\caption{{\small Comparison of the photometric redshifts from {\it Z-PEG} (model \ZPEG.4) with the HDF-N measured spectroscopic redshifts without J, H, K bands, and without any particular age constraint. These results have to be compared to Fig.~\ref{figure:fig8} for which there is no age constraint either. The dotted line is the linear regression of our photometric redshifts 
 if we exclude the points with $|\zZPEG-\zspe| \geq 1$.}}
\label{figure:fignoJHK}
\end{figure}
\end{center}

Since the strong discontinuities (4000\AA, Lyman break) are the most 
constraining features for redshift determinations, several consequences 
are implied (see also discussions in \cite{Bolzonella} and \cite{Massarotti}).
The observations of colors in the near infrared is the only way to follow
the discontinuities at the highest redshifts ($\le 4.5$ for the 4000\AA~discontinuity 
and $\le 23$ for the Lyman break in the K band). Moreover SED continua also contribute
to the best fits, so that in most cases, minimizing the $\chi^2$ on the largest
wavelength coverage will decrease the degeneracy.  Figure \ref{figure:fignoJHK} 
and table \ref{table:stats} show that without J, H and K bands, the photometric 
redshift determination is poor ($\sigmaz=0.32 > \sigmaz=0.12$ for $\zspe < 1.5$).
Moreover, we do not use any filter between $8500$ and $12000$ \AA~to fit the data,
because on the one hand we do not have the corresponding observations and on the other hand 
only very few galaxies in our sample are at $1<z<2$. 
To analyze galaxies in the redshift range $1<z<2$~, a $Z$ filter 
(or equivalent) would be necessary for observations as well as modeling. 
\subsection{IGM absorption}
When comparing results with and without IGM absorption (Fig.~\ref{figure:fig8}), 
an important deviation to the spectroscopic redshifts is observed
at high redshifts if we do not take the IGM into account: a linear regression 
on fits without the IGM shows a systematic deviation towards high photometric redshifts. This was already noted by 
Massarotti et al.(2001a,2001b), and can be understood quite easily: 
not including IGM absorption in models, one can mistake the Lyman forest 
absorption by intergalactic \ion{H}{i} (below $\lambda=1216$\AA) for 
the real Lyman break ($\lambda=912$\AA~in the rest frame). 
In such a case, we overestimate the redshift.
\section{Investigations on redshifts of formation}
\begin{center}
\begin{figure}[!htbf]
\begin{center}
\includegraphics{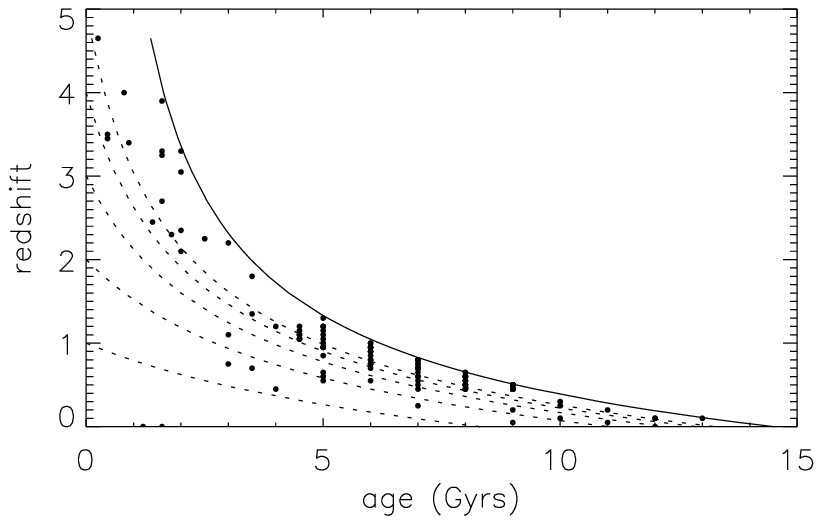}
\end{center}
\caption{{\small Galaxies of the HDF-N sample in the redshift--age diagram, after 
{\it Z-PEG} fits. The dotted lines show the $z$-age relations for various $\zfor$
(iso-$\zfor$ lines) and the standard cosmology, previously defined. For example, the 
dotted line crossing the y-axis (age=0) at $z=2$ corresponds to $\zfor=2$. The solid line 
shows the age of the universe at a given redshift (formally, this line is equivalent to
 $\zfor=\infty$). As a result, 
most galaxies form at $z>3$ (we can also check this result in Fig.~\ref{figure:zforhist}). }}
\label{figure:zfor}
\end{figure}
\end{center}

\begin{center}
\begin{figure}[!htbf]
\begin{center}
\includegraphics{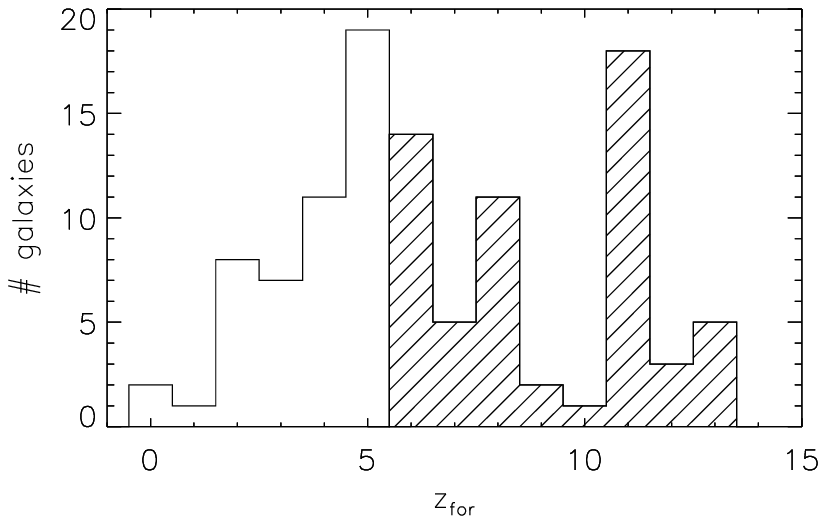}
\end{center}
\caption{{\small Histogram of $\zfor$ from our sample. For $\zfor > 5$, due to 
the lack of precision on the determination of (age, z$_{phot}$), we cannot conclude on $\zfor$ real values.  $\zfor= 5$ is then a lower value.}}
\label{figure:zforhist}
\end{figure}
\end{center}

Given the photometric redshift and age found with \ZPEG~for any observed galaxy, a 
corresponding formation redshift 
$z_{for}$~, defined as the epoch of the first star's appearance, may be
derived. The result, constrained by the age of the universe (\ageu),
depends on cosmology; as previously mentioned, standard cosmological parameters 
are hereafter adopted. 
Figure \ref{figure:zfor} shows iso-$z_{for}$~curves. The galaxy sample 
is also plotted and compared to curves. \\
For $\zfor<5$, our formation redshift 
determination is reliable, if we keep in mind that we constrained the ages 
of galaxies and thus the formation redshift. Unfortunately, the precision of redshift determination from photometry is quite 
poor (about $0.1$ for $z < 1.5$ with our sample). Above $\zfor=5$, the iso-$\zfor$ 
lines are so close to each other that we cannot rely on this formation redshift 
determination. \\
The histogram of the formation redshifts is shown in Fig.~\ref{figure:zforhist}.
Note that this histogram is in agreement with
\cite{Lanzetta2001} figures of the star formation rate as a function of redshift, deduced from 
\cite{FSLY}'s photometry, and taking into account the effects of cosmological surface brightness.
\section{Discussion}
The  determination of photometric  redshifts significantly  depends on
the adopted evolution scenarios. The basic principle is that distant  galaxies 
evolve according to the same basic principles as our nearby  galaxies, only  observed 
at  earlier epochs. As a direct consequence, evolution scenarios  must
reproduce the SEDs of galaxy templates observed at $z=0$. This constraint 
requires meaningful statistical  templates. The literature  appears
poor in this domain. From a large series of catalogs, we built statistical
samples corrected for various effects: aperture, inclination, reddening, etc
(see \cite{FRV99stats}) in order to 
focus specifically on visible to near-infrared colors of eight 
spectral types of galaxies. All our redshifted templates are 
compatible with these statistical colors at $z=0$.
  
The evolution principles are linked to the time-dependent star formation 
rates.  Corresponding to so-called "monolithic" scenarios, each type 
evolves with a star formation rate proportional to the gas content
and typical astration rates.  The time-scale of star 
formation increases from ellipticals (half Gyr) to Irregulars
(more than 10 Gyr). A test of the reliability of the adopted scenarios 
 is given with the interpretation of the deepest multi-spectral 
 faint galaxy counts on the largest multi-spectral coverage.
  Figure 5 of \cite{FRV99counts}
shows the best simultaneous fits of  the deepest surveys, including the 
HDF-N sample, from the far-UV to the K band. 
The most constraining data are the reddest and deepest counts, only fitted by scenarios of old
elliptical galaxy models. The evolution 
scenarios are from P\'EGASE and similar to those used in the code \ZPEG.
These considerations  make the $\zZPEG$ results
more robust and the evolution scenarios of elliptical galaxies 
compatible with a rapid evolution at the earliest
epochs. The high values of $z_{for}$ found from the present analysis
favor this result.
The star formation rate continuously follows the gas density, so that
the current SFR  depends on the past star formation history. 
An interesting interpretation, though tentative given the small incomplete
samples, would arise if the apparent continuous law is considered as the envelop
of small bursts. The intensity of these bursts decreases over time, 
as the gas is depleted by star formation. 

If the photometric redshifts  $\zZPEG$ are faintly dependent 
on the adopted cosmology, the results of  $z_{for}$~ are more sensitive to
the choice of the cosmological parameters through
the age-redshift relation. 

Finally the discussion of the dispersion and the degeneracy
of solutions has been detailed at each step of the $\zphot$~
determination. In this analysis, like the age constraint, 
the extension of the wavelength coverage, increasing with the number of filters,
limits the degeneracy. Conversely, if galaxies are observed 
through only a few filters, the degeneracy may be high and results
have to be used with caution. 
\section{Conclusions}
A new code \ZPEG~is proposed for public use, available on the web, 
to predict photometric redshifts from a data set of classical broad-band
colors or spectra. The code is derived from the model P\'EGASE, also available
on the same web site, frequently downloaded for evolution 
spectrophotometric analyses and quoted in many articles. 
The particularity of the analysis presented here is to underline the 
hypotheses implicitly (or explicitly) accepted when computing 
photometric redshifts with evolution. 
The comparison with a spectrophotometric redshift sample makes the predictions
robust. In its recent version P\'EGASE.2, the model
constrains typical evolution scenarios to fit nearby galaxies.
Type-dependent extinction, IMF
and metal effects are taken into account, although we assumed 
their modeling is not essential in our conclusions.
The most constraining parameter is the 
$z=0$~galaxy age, typical of the nearby stellar population of 
each spectral type. This constraint makes it possible to 
eliminate implausible secondary solutions, the degeneracy being 
the largest cause
of uncertainties on photometric redshifts. The dispersion for $z <1.5$
reaches its minimum value $0.0980$ with the age constraint, compared to other 
dispersion values whatever the other
parameter values (IGM absorption or NIR colors). 
The improvement in photometric redshift accuracy brought by near-infrared colors is
also measured by comparing models \ZPEG.4 and \ZPEG.2. The dispersion decreases from
$0.32$ to $0.12$ for $z < 1.5$ and is more important for high redshifts when the 4000\AA~discontinuity
enters in the NIR domain and the Lyman break does not yet reach the ultraviolet bands. 
Finally \ZPEG~is proposed to the community with P\'EGASE.2. The site will be 
updated and improved, so that \ZPEG~will also follow improvements in the spectrophotometric
evolution modeling, in particular with the series of flux-calibrated samples allowing better 
fits of highly redshifted templates. A possible extension of \ZPEG~will involve the adaptation 
of any photometric system, including narrow bands around emission lines,
with the help of the coupled code P\'EGASE+CLOUDY (\cite{Ferland}),
predicting stellar and photoionised gas emissions (\cite{Moy2001}).

\begin{acknowledgements}
We would like to thank Emmanuel Moy, Michel Fioc and Jeremy Blaizot for the fruitful discussions we had with them.
We are also grateful to the referee for his/her useful comments.
\end{acknowledgements}


\begin{thebibliography}{}

\bibitem[Arnouts et al.(1999)]{Arnouts1999} Arnouts, S., Cristiani, S., Moscardini, L., et al. 1999, \mnras, 310, 540

\bibitem[Baum(1962)]{Baum} Baum, W.~A.\ 1962, IAU Symp.~15: Problems of Extra-Galactic Research, 15, 390 

\bibitem[Ben{\' \i}tez(2000)]{Benitez} Ben{\' \i}tez, N.\ 2000, \apj, 536, 571 

\bibitem[Bolzonella et al.(2000)]{Bolzonella} Bolzonella, M., Miralles, J.\ -., \& Pell{\' o}, R.\ 2000,  \aap, 363, 476 

\bibitem[Budav{\' a}ri(2001)]{Budavari2001} Budavari, T.\ 2001, Publications of the Astronomy Deparment of the Eotvos Lorand University, 11, 41 

\bibitem[Budav{\' a}ri et al.(2001)]{Budavari} Budavari, T., Csabai, I., Szalay, et al.\ 2001, \aj, 122, 1163

\bibitem[Clegg \& Middlemass(1987)]{Clegg} Clegg, R.~E.~S.~\& Middlemass, D.\ 1987, \mnras, 228, 759 

\bibitem[Cohen et al.(2000)]{Cohen2000} Cohen, J.~G., Hogg, D.~W., Blandford, et al. 2000, \apj, 538, 29

\bibitem[Cohen(2001)]{Cohen2001} Cohen, J.~G.\ 2001, \aj, 121, 2895

\bibitem[Coleman et al.(1980)]{Coleman} Coleman, G.D., Wu, C.C., Weedman, D.W., 1980, Astrophys. J. Sup. Series, 43, 393

\bibitem[Connolly et al.(1995)]{Connolly} Connolly, A.~J., Csabai, I., Szalay, A.~S., et al. 1995, \aj, 110, 2655 

\bibitem[Cowie et al.(1994)]{Cowie} Cowie, L.~L., Gardner, J.~P., Hu, E.~M., et al. 1994, \apj, 434, 114 

\bibitem[Cowie et al.(1996)]{Cowie96} Cowie , L.~L., Songaila, A., Hu, E.~M., et al. 1996 \aj, 112, 839 

\bibitem[Dickinson(1998)]{Dickinson} Dickinson, M.\ 1998, The Hubble Deep Field, 219 

\bibitem[Dressler(1980)] {Dressler80} Dressler, A., 1980, \aj, 236, 351

\bibitem[FSLY]{FSLY} Fern{\' a}ndez-Soto, A., Lanzetta, K.\ M., \& Yahil, A.\ 1999, \apj, 513, 34 

\bibitem[Ferland(1996)]{Ferland} Ferland G. J., 1996. HAZY, a brief introduction to Cloudy, University of Kentucky, Department of Physics and Astronomy Internal Report

\bibitem[Fern{\' a}ndez-Soto et al.(2001a)]{FSLY2001a} Fern{\' a}ndez-Soto, A., Lanzetta, K.~M., Chen, H.-W., et al.\ 2001a, \apjs, 135, 41 

\bibitem[Fern{\' a}ndez-Soto et al.(2001b)]{FSLY2001b} Fern{\' a}ndez-Soto, A., Lanzetta, K.~M., Chen, H.-W., et al.\ 2001b, astro-ph/0111227

\bibitem[Fioc \& Rocca-$\!$Volmerange(1997)]{FRV1} Fioc, M.\ \& Rocca-$\!$Volmerange, B.\ 1997,  \aap, 326, 950 

\bibitem[Fioc \& Rocca-Volmerange(1999a)]{FRV99counts} Fioc, M.~\& Rocca-Volmerange, B. 1999a, \aap, 344, 393 

\bibitem[Fioc \& Rocca-$\!$Volmerange(1999b)]{FRV99stats} Fioc, M.~\& Rocca-$\!$Volmerange, B. 1999b, \aap, 351, 869 

\bibitem[Folkes et al.(1999)]{Folkes} Folkes, S., Ronen, S., Price, I., et al.\ 1999, \mnras, 308, 459

\bibitem[Fontana et al.(2000)]{Fontana-Arnouts} Fontana, A., D'Odorico, S., Poli, F., et al. \ 2000, \aj, 120, 2206 

\bibitem[Giallongo et al.(1998)]{Giallongo} Giallongo, E., D'Odorico, S., Fontana, A., et al.\ 1998, \aj, 115, 2169 

\bibitem[Koo(1985)]{Koo} Koo, D.~C.\ 1985, \aj, 90, 418 

\bibitem[Lanzetta et al.(2001)]{Lanzetta2001} Lanzetta, K., Yahata, N., Pascarelle, S., et al., astro-ph/0111129

\bibitem[Le F\`evre(1995)] {CFRS} Le F\`evre, O., Crampton D., Hammer F., Tresse L., 1995, \apj, 455, 60

\bibitem[Leitherer et al.(1996)]{Crete} Leitherer, C. ,Fritze-von Alvensleben, U., Huchra, J., eds., in Proceedings of "From Stars to Galaxies: The Impact of Stellar Physics on Galaxy Evolution", ASP Conf. Series Vol. 98, Crete 1995. 

\bibitem[Lejeune, Cuisinier, \& Buser(1997)]{Lejeuneetali} Lejeune, T., Cuisinier, F., \& Buser, R.\ 1997, \aaps, 125, 229

\bibitem[Lejeune et al.(1997, 1998)]{Lejeuneetal} Lejeune, T., Cuisinier, F., \& Buser, R.\ 1998, \aaps, 130, 65

\bibitem[Loh \& Spillar(1986)]{Loh} Loh, E.~D.~\& Spillar, E.~J.\ 1986, \apj, 303, 154 

\bibitem[Madau(1995)]{Madau95} Madau, P.\ 1995, \apj, 441, 18

\bibitem[Madau et al.(1996)]{Madau96} Madau, P., Ferguson, H.\ C., Dickinson, et al.\ 1996, \mnras, 283, 1388 

\bibitem[Massarotti et al.(2001a)]{Massarotti} Massarotti, M., Iovino, A., \& Buzzoni, A.\ 2001a,  \aap, 368, 74 

\bibitem[Massarotti et al.(2001b)]{Massarottib} Massarotti, M., Iovino, A., Buzzoni, A., et al.\ 2001b, \aap, 380, 425

\bibitem[Moy et al.(2001)]{Moy2001} Moy, E., Rocca-Volmerange, B., \& Fioc, M.\ 2001, \aap, 365, 347

\bibitem[Press et al.(1993)]{Press93} Press, W.~H., Rybicki, G.~B., \& Schneider, D.~P.\ 1993, \apj, 414, 64

\bibitem[Rana \& Basu(1992)]{Rana} Rana, N.\ C.\ \& Basu, S.\ 1992, \aap, 265, 499 

\bibitem[Rocca-Volmerange(2000)]{Rocca00} Rocca-Volmerange, B., 2000, in Galaxy Disks and disk galaxies, ed Fun\'es \& Corsini, PASP Series, 230, 597

\bibitem[Steidel et al.(1999)]{Steidel} Steidel, C.~C., Adelberger, K.~L., Giavalisco, M., et al.\ 1999, \apj, 519, 1 

\bibitem[Williams et al.(1996)]{Williams} Williams, R.~E., Blacker, B., Dickinson, M., et al.\ 1996, \aj, 112, 1335

\bibitem[Woosley \& Weaver(1995)]{WW} Woosley, S.~E.~\& Weaver, T.~A.\ 1995, \apjs, 101, 181 

\end{thebibliography}
\end{document}